\documentclass[fleqn,useAMS,longbibliography]{nrc1}
\usepackage{nrcdoc}
\usepackage{graphicx}
\usepackage{amsmath}
\usepackage{amssymb}
\usepackage{listings}
\usepackage[square, sort, numbers,sort&compress]{natbib}
\usepackage{lipsum}
\usepackage{multicol}
\usepackage{bm}
\usepackage[english]{babel}
\usepackage{footmisc}
\title{The Phantom of RAMSES user guide for galaxy simulations using Milgromian and Newtonian gravity}

\author[Srikanth Togere Nagesh]{S. T. Nagesh}
\correspond{tnsrikanth@uni-bonn.de}
\address[AIFA]{Argelander-Institut f\"ur Astronomie, Universit\"at Bonn, Auf dem H\"ugel 71, 53121 Bonn, Germany.}
\author[Indranil Banik]{I. Banik}
\address[HISKP]{Helmholtz-Institut f\"ur Strahlen- und Kernphysik, Universit\"at Bonn,
Nussallee 14-16, 53115 Bonn, Germany.}
\author[Ingo Thies]{I. Thies}
\address[HISKP]
\author[Pavel Kroupa]{P. Kroupa}
\address{Helmholtz-Institut f\"ur Strahlen- und Kernphysik, Universit\"at Bonn,
Nussallee 14-16, 53115 Bonn, Germany and Astronomical Institute, Faculty of Mathematics and Physics; Charles University in Prague, V Hole\v{s}ovi\v{c}k\'ach 2, CZ-180 00 Praha, Czech Republic.}
\author[Benoit Famaey]{B. Famaey}
\address[OBAS]{Universit\'{e} de Strasbourg, CNRS UMR 7550, Observatoire astronomique de Strasbourg, 11 rue de l'Universit\'{e}, 67000 Strasbourg, France}
\author[Nils Wittenburg]{N. Wittenburg}
\address[HISKP]
\author[Rachel Parziale]{R. Parziale}
\address[AIFA]
\author[Moritz Haslbauer]{M. Haslbauer}
\address[MPIFR]{Max-Planck-Institut f\"ur Radioastronomie, Auf dem H\"ugel 69, 53121 Bonn, Germany and Helmholtz-Institut f\"ur Strahlen- und Kernphysik, Universit\"at Bonn, Nussallee 14-16, 53115 Bonn, Germany.}

\begin{document}

\maketitle

\begin{abstract}

This document describes the general process of setting up, running, and analysing disc galaxy simulations using the freely available program \textsc{phantom of ramses} (\textsc{por}). This implements Milgromian Dynamics (MOND) with a patch to the \textsc{ramses} grid-based $N$-body and hydrodynamical code that uses adaptive mesh refinement. We discuss the procedure of setting up isolated and interacting disc galaxy initial conditions for \textsc{por}, running the simulations, and analysing the results. This manual also concisely documents all previously developed MOND simulation codes and the results obtained with them.

\end{abstract}

\begin{multicols}{2}

\section{Introduction}

Milgromian Dynamics (MOND) is an extension of Newtonian dynamics to encompass the observed dynamics in the Solar System as well as in galaxies without postulating invisible haloes around them \citep{Milgrom_1983}. MOND computes the gravitational potential of galaxies using only the distribution of baryons. It has been very successful in this regard, especially because it predicted some very tight scaling relations which were subsequently observed \citep{Famaey_McGaugh_2012, Lelli_2017}. These are a consequence of Milgrom's formula
\begin{eqnarray}
    g ~=~ \sqrt{g_\mathrm{N} a_{_0}}  ~~~\textrm{for} ~~ g_\mathrm{N} \ll a_{_0} = 1.2 \times 10^{-10} \, \textrm{m s}^{-2},
\end{eqnarray}
where $a_{_0}$ is Milgrom's constant, $g$ is the strength of the true gravity, and $g_\mathrm{N}$ is that of the Newtonian gravity. To achieve a generalization of gravity applicable in non-spherical systems, MOND requires a generalized Poisson equation derived from a Lagrangian. Two classical variants have been proposed, one with an aquadratic Lagrangian \citep[AQUAL;][]{Bekenstein_Milgrom_1984}, and one with a Lagrangian making use of an auxiliary field, which is called the quasi-linear formulation of MOND \citep[QUMOND;][]{QUMOND}. MOND may be a consequence of the quantum vacuum \cite{Milgrom_1999, Pazy_2013, Verlinde_2016, Smolin_2017}. Reviews of MOND can be found in \citep{Famaey_McGaugh_2012, Milgrom_2015_review}.

\textsc{phantom of ramses} \citep[\textsc{por};][]{Lughausen_2015} is a numerical implementation of QUMOND, whose field equation for the potential $\Phi$ is
\begin{eqnarray}
    \nabla^2 \Phi ~\equiv~ - \nabla \cdot \bm{g} ~=~ - \nabla \cdot \left( \nu \bm{g}_\mathrm{N} \right) \, ,
    \label{eq:poisson}
\end{eqnarray}
where $\nu$ is the MOND interpolating function with argument $y \equiv g_\mathrm{N}/a_{_0}$, with $\bm{g}$ and $\bm{g}_\mathrm{N}$ being the true and Newtonian gravitational acceleration vectors, respectively, and $v \equiv \left| \bm{v} \right|$ for any vector $\bm{v}$. The current version of \textsc{por} uses the simple form of the interpolating function (e.g. equation 5 of \citep{
Banik_2018_Centauri})
\begin{eqnarray}
    \nu \left( y \right) ~=~ \frac{1}{2} + \sqrt{\frac{1}{4} + \frac{1}{y}} \, .
    \label{eq:nu}
\end{eqnarray}
$\bm{g}_\mathrm{N}$ is found from the baryonic density $\rho_b$ using the standard Poisson equation
\begin{eqnarray}
    \nabla \cdot \bm{g}_\mathrm{N} ~=~ -4 \mathrm{\pi} G \rho_b \, .
    \label{eq:poisson_Newton}
\end{eqnarray}
The boundary condition for the MOND potential far from an isolated matter distribution is
\begin{eqnarray}
    \Phi ~=~ \sqrt{GMa_{_0}} \ln R \, ,
    \label{eq:boundary_MOND}
\end{eqnarray}
where $M$ is the mass within the simulation box, and $R$ is the distance from its barycentre in the simulation unit of length.

A handful of Milgromian $N$-body codes were developed before \textsc{por} to handle MOND computations, and these have been applied to various problems. The first multi-grid, Milgromian $N$-body code was developed by \citep{Brada_1999} to investigate the stability of disc galaxies. This was later extended to simulate how they might warp due to the external field effect \citep[EFE;][]{Brada_2000}. Another $N$-body solver which implemented the AQUAL formulation of MOND was developed and used to study the evolution of spiral galaxies using pure stellar discs \citep{Tiret_2007}. Gas dynamics was later included using a sticky particle scheme at low resolution \citep{Tiret_2008_gas}. \textsc{n-mody} was developed to solve the Milgromian Poisson equation in spherical coordinates \citep{Londrillo_2009} and used to investigate dynamical friction \citep{Nipoti_2008}, orbit instabilities \citep{Nipoti_2011}, and stellar kinematics \citep{Wu_2013, Wu_2018, Wu_2019}. Milgromian $N$-body codes tailored to cosmological simulations have also been developed \citep{Llinares_2008, Llinares_2011, Angus_2011, Angus_2013}. Another $N$-body solver called \textsc{raymond} \citep{Candlish_2015} was developed to implement both the AQUAL \citep{Bekenstein_Milgrom_1984} and QUMOND \citep{QUMOND} formulations of MOND. \textsc{raymond} has been applied to cosmological \citep{Candlish_2016}, galaxy cluster \citep{Candlish_2018}, and other problems. However, not all the aforementioned $N$-body codes can be applied to generic scenarios simultaneously involving particles, gas dynamics, and star formation.

The \textsc{fortran}-based \textsc{por} code was developed by Fabian L\"ughausen \citep{Lughausen_2015}. It is a customized version of \textsc{ramses} \citep{Teyssier_2002}, which exclusively uses Newtonian dynamics to compute gravity. \textsc{por} can compute gravity using MOND by numerically solving Equation \ref{eq:poisson}. Since \textsc{por} is a patch to \textsc{ramses}, it inherits use of the adaptive mesh refinement (AMR) technique. \textsc{por} is equipped to handle particles, gas dynamics, and star formation, and can be applied to diverse problems. It allows the user to compute gravity in both Milgromian and Newtonian frameworks \citep{Lughausen_2015}.

This document serves as a tutorial/manual for the general use of \textsc{por}, with some suggestions for the specific case of setting up and simulating a disc galaxy. Most of the steps and parameters described here are specific to \textsc{por}, except the installation of \textsc{ramses}. For a detailed description of individual parameters, it is always recommended to read the \textsc{ramses} manual\footnote{\url{https://bitbucket.org/rteyssie/ramses/src/master/}\label{RAMSES}}. Most of the parameters and files described here can be edited safely without disturbing the core algorithms. Before changing parameters or files that are not mentioned here, it is important to fully understand the workings and consequences of the change.

In Section \ref{sec:Installation}, we explain the installation procedure of \textsc{ramses} and \textsc{por}. In Section \ref{sec:dice}, we explain how to set up MOND disc templates using Disk Initial Conditions Environment (\textsc{dice}), and thereby generate rotating disc initial conditions for \textsc{por}. In Section \ref{sec:running}, we describe the workings of \textsc{por}, focusing on particle-only and hydrodynamical runs with and without star formation. In Section \ref{sec:random}, random turbulence generation is briefly discussed. Section \ref{sec:extraction} discusses the \textit{extract\_por} tool used to analyse particle data in \textsc{ramses} simulation outputs. In Section \ref{sec:Test}, we mention all publications based on \textsc{por}. We conclude in Section \ref{Conclusions}.

\section{Installation and setup of the code}
\label{sec:Installation}

The \textsc{por} patch by Fabian L\"ughausen is rated to work with the 2015 
version of \textsc{ramses}, which has since been modified (the latest \textsc{ramses} version is available here \textsuperscript{\ref{RAMSES}}). Later versions are not compatible with \textsc{por}. It is therefore recommended to use the 2015 version of \textsc{ramses} with \textsc{por}. The jointly tested version of \textsc{ramses} and \textsc{por} is available here \footnote{\label{PoRbit}\url{https://bitbucket.org/SrikanthTN/bonnpor/src/master/}}, in the \textit{PoR\_hydro} folder.

The following steps describe the installation and compilation of \textsc{ramses} and \textsc{por}. These procedures are adapted from the \textsc{ramses} manual, where they are described further. 

\begin{enumerate}
    \item The main folder needed for compilation of \textsc{ramses} is \textit{bin}. In the \textit{bin} folder, there is a \textit{makefile}. Now, do:
    
\begin{lstlisting}[language=bash]
$ cd ~/PoR_hydro/ramses/bin/
\end{lstlisting}

	\item In the \textit{makefile}, certain flags need to be changed.
	\textit{Makefile}:
	
	\hspace{-0.3in}\fbox{\begin{minipage}{25em}
	
	Compilation time parameters\\
	NVECTOR = 32\\
	NDIM = 3\\
	NPRE = 8\\
	NVAR = 6\\
	NENER = 0\\
	SOLVER = hydro\\
	\#PATCH = ../patch/phantom\_units \\
	\#PATCH = ../patch/phantom\_staticparts ~~~~ (particle-only run)\\
	\#PATCH = ../patch/hydro/phantom\_merger \\
	PATCH = ../patch/hydro/phantom\_extfield ~~ (hydro run) \\
	EXEC = RAMSES
	\end{minipage}}

	\item All these flags are explained in the \textsc{ramses} manual\textsuperscript{\ref{RAMSES}}. \textit{F90} and \textit{FFLAGS} should be set carefully.

	\hspace{-0.3in}\fbox{\begin{minipage}{25em}
	
	F90 = mpif90 -frecord-marker=4 -O3 -ffree-line-length-none -g -fbacktrace \\
	FFLAGS = -x f95-cpp-input \$(DEFINES)\$
	
	\end{minipage}}

	\item  \textit{F90} sets the Fortran compiler and \textit{FFLAGS} is used to specify the required MPI libraries, which are mainly used for parallel computing. This is important given the likely high computational cost. The default \textit{makefile}\textsuperscript{\ref{PoRbit}} uses the above-mentioned \textit{F90} and \textit{FFLAGS}. If one's computer is not compatible with these default parameters, they can be changed in the \textit{makefile}.

	\item Once all the required flags are set, compile the code:
\begin{lstlisting}[language=bash]
$ make 
\end{lstlisting}

	After compilation, one can test the installation as described in section 2.3 of the \textsc{ramses} manual\textsuperscript{\ref{RAMSES}}. 
	
	\item To make the files again, go to the \textit{bin} folder and execute:
	
\begin{lstlisting}[language=bash]
$ make clean
$ make
\end{lstlisting}
	
	 The \textsc{ramses} manual was written in 2002 and has not been updated since, so there might be subsequent modifications to the parameter file. One must use the \textit{phantom} patch to do simulations in MOND. For Newtonian simulations, it is recommended to use this patch and set the mond flag to \textit{.false.} in the \textit{namelist}.
\end{enumerate}

\subsection{Compilation of the code with the \textsc{por} patch}
\label{sec:compilation}

We now describe the procedure to link the \textsc{por} patch and re-compile \textsc{ramses}. To activate the \textsc{por} patch, the following parameters must be specified in the \textit{makefile} (\# means a comment):

\hspace{-0.3in}\fbox{\begin{minipage}{26em}
 NDIM = 3 \\
 PATCH = ../patch/phantom\_staticparts  \\
 \#PATCH = ../patch/phantom \\
 \#PATCH = ../patch/hydro/phantom\_extfield
\end{minipage}} \\

By default, \textsc{ramses} uses periodic boundary conditions, but the \textsc{por} patch in\textsuperscript{{\ref{PoRbit}}} specifies and uses different boundary conditions appropriate to isolated galaxy simulations. The 2015 version of \textsc{ramses} available in\textsuperscript{\ref{PoRbit}} contains a staticpart patch and a hydrodynamical patch with compulsory additional merger and EFE patches whose effects can be disabled (Section \ref{sec:hydronosfr}). These patches are task-specific customizations of \textsc{por}. For a particle run, \textit{phantom\_staticparts} can be used while for a hydrodynamical run, \textit{phantom\_extfield} can be used. Only one patch can be used at a time. One must change the path to the user's directory before making the file. After specifying the parameters, make the file again.

\section{Disk Initial Conditions Environment (\bf \textsc{DICE})}

\label{sec:dice}
Any galaxy or cosmological simulation needs initial conditions. The \textsc{ramses} user guide refers to two websites for these, but they are for cosmological runs. The \textsc{music}\footnote{\label{MUSICcode}\url{https://bitbucket.org/ohahn/music/src/master/}} code also provides initial conditions for the latter, and is recommended. The setup of cosmological MOND simulations will be discussed elsewhere. This guide focuses on galaxy simulations, for which we generate initial conditions with an adapted version of Disk Initial Conditions Environment \citep[\textsc{dice};][]{Perret_2014}. The original version of \textsc{dice} can be found here\footnote{\label{DICE1}\url{https://bitbucket.org/vperret/dice/src/master/}}. It is not compatible with MOND or the 2015 version of \textsc{ramses}\textsuperscript{\ref{PoRbit}}, so we used a modified version of \textsc{dice} available here\textsuperscript{\ref{PoRbit}}. This has two versions, one for particle-only runs (the \textit{dice\_particle} folder, hereafter \textsc{p-dice}) and the other for hydrodynamical runs (in \textit{dice\_gas}, hereafter \textsc{h-dice}). These algorithms were developed by Graeme Candlish, Roy Truelove, Indranil Banik, and Ingo Thies. Both are equipped to initialize disc galaxies in MOND, but in principle other methods could be used and advanced with the \textsc{por} patch. Before installing \textsc{dice}, \textit{CMake}, \textit{GSL}, and \textit{FFTW} must be installed. If this is not already the case, installation instructions are provided here\textsuperscript{\ref{PoRbit}}.

\subsection{Installation and setup}
\label{sec:instalDICE}

As mentioned above, the folders \textit{dice\_particle} and \textit{dice\_gas} contain \textsc{p-dice} and \textsc{h-dice}, respectively. Extract them to \textbf{/local} in one's home directory. In \textit{dice\_particle}, the \textit{disc} folder is required for disc galaxy simulations. Now, in \textit{disc}, the \textit{bin} folder contains the \textit{makefile} needed for compilation, while the \textit{example} folder contains the parameter files. To compile \textsc{p-dice}, execute:

\begin{lstlisting}[language=bash]
$ cd dice_particle 
$ cd disc
$ mkdir build 
$ cd build
$ cmake ..
$ make
$ make install
\end{lstlisting}

To compile \textsc{h-dice}, execute:

\begin{lstlisting}[language=bash]
$ cd dice_gas
$ mkdir build 
$ cd build
$ cmake ..
$ make
$ make install
\end{lstlisting}

\textsc{h-dice} does not contain an additional \textit{disc} folder like \textsc{p-dice}.

\subsection{Running DICE}
\label{sec:runningDICE}

The \textit{dice\_gas} and \textit{disc} (in \textit{dice\_particle}) folders contain four sub-folders:
\begin{enumerate}
	\item \textit{cmake} should not be altered,
	\item \textit{build} will contain the executable,
	\item \textit{src} contains the source files which encode the physics required for computation, and
	\item \textit{example} contains files required to generate the initial conditions, with task-specific configuration files like \textit{M31, M33}, and generic scenarios like a disc galaxy, disc with a bulge etc. Only the Milky Way (MW), M31, and M33 cases are rated to work.
\end{enumerate}

We used the \textit{test\_mw.config} configuration file for our disc galaxy:
\\
\\
\hspace{-0.2in}\fbox{\begin{minipage}{25em}
Redshift 	~~3.0 \\
Galaxy 		~~../../example/params\_files/testMilkyWay.params \\    
Filename 	~~dice\_highz \\
ICformat 	~~Gadget2 \\
Nthreads 	~~32
\end{minipage}}
\\

In the \textit{.config} file, specify the path to the parameter file. The redshift is unused in our MONDified \textsc{dice}. The \textit{testMilkyWay.params} is the parameter file used, though other \textit{params} files exist in the \textbf{/example/params\_files} folder. Custom templates can be created using these parameter files, though only the MW, M31, and M33 cases are rated to work. There are mainly three types of parameters in \textit{testMilkyWay.params}: \textit{global parameters}, \textit{outer disc}, and \textit{inner disc}. For both \textsc{p-dice} and \textsc{h-dice}, once the parameter file is set, go one directory up and execute:
\begin{lstlisting}[language=bash]
$ cd bin
$ ./dice ../../example/test_mw.config
\end{lstlisting}
After execution, 2 output files named \textit{Milky\_Way\_output\_p2\_k0.txt} and \textit{Milky\_Way\_rotation\_curve.txt} will be created in the \textit{bin} folder. The rotation curve is only required for hydrodynamical simulations.

\section{Running \bf\textsc{por}}
\label{sec:running}

After compilation of \textsc{ramses} with the required \textsc{por} patch, one can customize the \textit{namelist} file available in the \textit{PoR\_namelist} folder to meet a scientific goal. \textit{PoR\_namelist} consists of all the \textit{namelist} files we have used for our runs. \textit{PoR.nml} is a general template which can be customized. \textit{PoR-static.nml} is the file we used for our particle-only run, while \textit{Test\_hydro\_mw\_NSFR.nml} was used for the hydrodynamical run without star formation. There is a general \textit{namelist} folder which consists of \textit{.nml} files that can be used to test e.g. the installation of \textsc{ramses}.

\subsection{Particle-only run (staticpart patch)}
\label{sec:static}

Use the \textit{/patch/phantom\_staticparts} patch and \textit{PoR-static.nml}, take care on the boundary conditions:
\\
\\
\fbox{\begin{minipage}{25em}
\&RUN\_PARAMS \\
poisson=.true. \\
pic=.true. \\
mond=.true.     --    Activates MOND poisson solver\\
nrestart=0   -- used to restart a run from any output\\
/ \\

\&AMR\_PARAMS  . \\
levelmin=7 \\
levelmax=12 \\
ngridmax=2000000   \\
boxlen=1024.0 \\ 
npartmax=2000000 
/ \\

\textit{ngridmax} and \textit{npartmax} should be of order $10^6$ to avoid memory errors.
\\

\&OUTPUT\_PARAMS \\
foutput=8000 -- Frequency of outputs in terms of coarse steps \\
noutput=100   -- Number of outputs to be generated \\
delta\_tout=100. -- Interval of the output in Myr \\
tend=10000. -- Simulation end time, in this case 10 Gyr \\
/ \\

\&INIT\_PARAMS \\
filetype=`ascii' \\
initfile= ../path/to/\textit{Milky\_Way\_output\_p2\_k0.txt} as the input. \\
/

\&POISSON\_PARAMS \\
a0\_ms2=1.2e-10 \\ 
m\_threshold=1.e+30  -- critical part, set it based on usage\\
gravity\_type=0 \\
cg\_levelmin=999 \\
/

\&BOUNDARY\_PARAMS \\
nboundary=6  		\\
ibound\_min=-1, 1, 0, 0, 0, 0, \\
ibound\_max=-1, 1, 0, 0, 0, 0, \\
jbound\_min= 0, 0,-1, 1, 0, 0, \\
jbound\_max= 0, 0,-1, 1, 0, 0, \\
\end{minipage}}
\fbox{\begin{minipage}{25em}
kbound\_min= 0, 0, 0, 0,-1, 1, \\
kbound\_max= 0, 0, 0, 0,-1, 1, \\
bound\_type= 1, 1, 1, 1, 1, 1, \\
/
\end{minipage}}
\\

There are some parameters which are mandatory in all runs, such as \textbf{\&Run\_Params}, \textbf{\&AMR\_Params}, \textbf{\&Output\_Params}, \textbf{\&Init\_Params}. Others vary based on specific requirements. Most of these parameters are detailed in the \textsc{ramses} manual\textsuperscript{\ref{RAMSES}}, so we only stress those specific to \textsc{por}. Parameters not shown here should not be changed unless required.

The staticpart patch integrates particles below a certain mass \textit{m\_threshold}, while more massive particles are kept static but are considered when evaluating $\bm{g}_\mathrm{N}$ in Equation \ref{eq:poisson_Newton}. This method is an effective way to save computation time. If one wants to evolve all the stellar particles in a particle-only simulation, then \textit{m\_threshold} should be set to a suitably large value, e.g. $10^{30} M_\odot$. The units from the \textsc{dice} output are the same as required by \textsc{por} for input (i.e. $M_\odot$, kpc, and km/s), while \textit{units.f90} has the units used by \textsc{por} in which $G = 1$.

One can modify \textbf{\&Output\_Params} and \textbf{\&AMR\_Params}, but it is not recommend to tamper with other blocks. In the above example, the \textit{Milky\_Way\_output\_p2\_k0.txt} obtained from \textsc{p-dice} is given as the input file, with the rotation curve unused. Once all parameters are set in the \textit{namelist} file, the simulation can be started by executing:

\begin{lstlisting}[language=bash]
$ mpiexec -n 32 ../ramses3d  ../filename.nml  
\end{lstlisting}

This calls the simulation to run on 32 CPUs using parallel computing (the number can be changed). Regardless of the directory of execution, one must specify full paths to the \textit{ramses3d} and \textit{namelist} files. To run these simulations without parallel computing, simply execute:

\begin{lstlisting}[language=bash]
$ ../bin/ramses3d  ../filename.nml
\end{lstlisting}
Users should check the computing capacity before running simulations without parallel computing.

After starting the simulation, it might terminate with error message \textit{- ``SEGSEV - invalid memory reference''}. This is a memory error, which can be solved by increasing \textit{npartmax} up to $10^7$ and \textit{ngridmax} up to $8 \times 10^6$ (the codes are not rated for larger values). The \textit{npartmax} variable must be at least equal to the number of particles in the \textsc{dice} template. Turning off the \textit{movie} may help. CPU and memory errors are a bit alarming, but are easily overcome and should not be a big concern for beginners.

The simulation will produce output folders for each snapshot. During the run, if the memory allocated is too small, the simulation will stop and ask to increase the number of grid cells. One must then go back to the \textit{namelist} file and increase \textit{ngridmax} or \textit{npartmax} based on what is asked. The restart protocol is rated to work, so restart the run from the last output file by setting \textit{nrestart} to the desired output number (the default of 0 means to start from scratch). If the run stops before finalising \textbf{output\_..45}, set \textit{nrestart = 45} and resume the run by executing the above-mentioned command.

\subsection{Hydrodynamical run without star formation}
\label{sec:hydronosfr}

We performed this run with the \textit{/patch/hydro/phantom\_extfield} patch, which is a modification to \textsc{por}. The EFE and merger scenarios are included using the MOND Poisson solver, but both features can be turned off.

\subsubsection{DICE with gas component}

\textsc{dice} is again used to set up the initial conditions. Since the gas component is included, we used \textsc{h-dice} in the \textit{dice\_gas} folder. To include a gas component, the \textit{test\_MilkyWay.params} was slightly modified:
\\

\hspace{-0.3in}\fbox{\begin{minipage}{26em}

\#\#\#\#\#\#\#\#\#\#\#\#\#\#\#\#\#\# \\
\# Global parameters \\
\#\#\#\#\#\#\#\#\#\#\#\#\#\#\#\#\#\#\# \\ 
\# Virial velocity of the galaxy [km/s] \\
v200 			~~	200.0 \\ 
\# Virial mass of the galaxy [1e10 Msol] \\
\# Overrides the v200 parameter \\ 
m200			~~	9.15\#8.4 old \\
\textbf{Gas\_fraction ~~0.2} \\ 
\textbf{Gas\_T ~~   50000.0}
\end{minipage}}
\\

The highlighted lines are the new additions to the \textsc{h-dice} template. These lines specify the gas component parameters. The gas fraction depends on the galaxy, here 20\% gas fraction was used for the MW. The gas temperature should be set equal to another parameter called \textit{T2\_ISM}, which is present in the \textit{namelist} file of \textsc{por}. One must be careful that the gas fraction should be greater than the mass fraction of the outer component, in this particular case, more than 18\%. This is because the distribution of gas in \textsc{h-dice} is done in a particular way, so one should be cautious while setting the gas fraction \citep{Banik_2020_M33}. The template has a default mass fraction of 17.64\% for the outer disc component, with the remaining 82.36\% for the inner disc \citep{Banik_2018_escape}. This version is only rated for two exponential disc components. For a beginner, it is recommend to take advice at this point before proceeding further.

\textsc{h-dice} can be started the same way as \textsc{p-dice}. After starting the \textsc{dice} run, one might notice a message in the terminal {\small ``\textbf{{Gas is too cold to satisfy the Toomre condition everywhere. Increase $T$ by a factor of \ldots}}''}, and/or {\small{``\textbf{WARNING: only writing data for component 1}}}''. The first message is just a warning about the global disc stability, and has no impact on the results $-$ it can be ignored. Temperature here is used as a measure of velocity dispersion including turbulence, so it is not the true gas temperature. The second message can also be ignored $-$ it indicates that the second component defined in \textsc{dice} is treated as stars and used for calculating the potential, but not printed in the output as the gas will be added in \textsc{por} \citep{Banik_2020_M33}. The particle data written to the disc template file contains only the stellar component.

The rotation curve file has columns for the gas disc scale height and its radial gradient. This is critical as the gas component is created in \textsc{por} itself (in the \textit{merger} and \textit{extfield} patch) by reading in the gas data from the rotation curve file and some parameters to be set in the \textit{namelist} file, e.g. gas mass and temperature. Care is needed to ensure compatibility of the parameters used for \textsc{dice} and \textsc{por}.

\subsubsection{\textsc{por} with merger and external field patch}
\label{sec:PoRmerger}

In the hydrodynamical case, we did not explicitly set up an isolated disc galaxy, but instead adapted the merger template \textit{condinit}\footnote{\url{http://www.physics.usyd.edu.au/~tepper/codes/ramses/trunk/doc/html/patch_2hydro_2merger_2condinit_8f90_source.html}\label{Chapon_2010}}. This sets up two disc galaxies in the simulation box, so we switched off the second galaxy by setting its mass and velocity to zero and placing it outside the simulation box. The \textit{namelist} file for each run should be customized as required, we show part of \textit{Test\_hydro\_mw\_NSFR.nml} as an example:
\\

\hspace{-0.2in}\fbox{\begin{minipage}{25em}
\&RUN\_PARAMS \\
mond=.true. \\ 
Activate\_g\_ext=.false. \\

\&INIT\_PARAMS \\
filetype=`ascii' \\
initfile(1)=`path/to/the/DICE/output/' \\
\

\&MERGER\_PARAMS \\
rad\_profile=`double\_exp' \\
z\_profile=`sech\_sq' \\
Mgas\_disc1=45.75 \\
Mgas\_disc2=0 \\
IG\_density\_factor=1.0e-2 \\
T2\_ISM=40.d3 \\
scale\_a2=1. \\
Vcirc\_dat\_file1=`Milky\_Way\_rotation\_curve.txt' \\
Vcirc\_dat\_file2=`Milky\_Way\_rotation\_curve.txt'  \\
ic\_part\_file\_gal1=`Milky\_Way\_output\_p2\_k0.txt'   \\
ic\_part\_file\_gal2=`Milky\_Way\_output\_p2\_k0.txt' \\
gal\_center1= 0.,0.,0. \\ 
gal\_center2= 2000,0.,0. \\
Vgal1=0.,0.,0. \\
Vgal2=0.,0.,0.
\end{minipage}}
\\

The \textit{namelist} has other parameters, but we show only those critical to the simulation. If one were to use a similar setup, the following suggestions are helpful:

\begin{enumerate}

    \item In the \textsc{ramses} \textit{makefile}, one must provide the path to the external field patch and recompile \textsc{ramses}. 
    
    \item The \textsc{por} patch can accommodate both MOND and Newtonian physics. The latter is used if one sets \textit{mond = .false.}, allowing simulations with both gravity theories using \textsc{por}.

    \item Setting \textit{Activate\_g\_ext} to false turns the EFE off. Hydrodynamical simulations with the EFE are discussed further in \citep{Banik_2020_M33}.


    \item For the \textit{initfile(1)}, one must give the path to the directory where \textit{Milky\_Way\_output\_p2\_k0.txt} and \textit{Milky\_Way\_rotation\_curve.txt} are present. These files should be specified for \textit{ic\_part\_file\_gal1} and \textit{Vcirc\_dat\_file1}, respectively. The same path and files can be given for the second galaxy, which is unused here. 

	\item The main things that need attention are the {\small{\textbf{\&Merger\_Params}}}. The gas mass of the galaxy is in units of $10^9 M_{\odot}$. If simulating interacting galaxies, they should not start too close together. We switched the second galaxy off by setting its gas mass and velocity to 0 and placing it outside the box, e.g. box size = 500 kpc, gal\_center2 = (2000, 0, 0) kpc. The first galaxy was placed at the box centre.

	\item For isolated simulations, both galaxies should have zero initial velocity, i.e \textit{Vgal1} and \textit{Vgal2} should be zero. \textit{gal\_axis} defines the disc's spin axis. For standard isolated simulations, use $\left( 0, 0, 1 \right)$ for counter-clockwise rotation around the $z$-axis, or $\left( 0, 0, -1 \right)$ for clockwise.

	\item The \textit{T2\_ISM} parameter in the \textit{namelist} and \textit{Gas\_T} in the \textsc{h-dice} template should be equal. The temperature floor \textit{T2\_star} should be set to a slightly lower value than \textit{T2\_ISM}. We used \textit{T2\_ISM} = 40,000 K and \textit{T2\_Star} = 30,000 K. The simulation is not rated to work with \textit{T2\_ISM} or \textit{T2\_Star} below 25,000 K.
   
\end{enumerate}

After taking care of all these parameters, one can start the run, leading to creation of the output folders in due course. The simulations are RAM and memory intensive $-$ a hydrodynamical disc galaxy advanced for 1.5 Gyr on a 4-core laptop could take a week or two depending on the RAM and might occupy up to 100 GB of hard disk space. These estimates would vary depending on the parameters and machine used $-$ see section 4.2 of the \textsc{ramses} manual\textsuperscript{\ref{RAMSES}} for more details.

\subsection{Hydrodynamical run with star formation}

Converting gas into stars requires careful treatment of baryons, for which \textsc{ramses} is well equipped and tested \citep{Rasera_2006}. Since \textsc{por} is just a modification to the Poisson solver, it does not affect the baryon treatment, inheriting that of standard \textsc{ramses}.

To activate star formation, one has to include the {\small{\textbf{\&Physics\_Params}}} in the \textit{namelist} file. One can add {\small{\textbf{\&Physics\_Params}}} to the \textit{Test\_hydro\_mw\_NSFR.nml} and activate star formation. Alternatively, one can use the \textit{MW\_hydro\_SFR.nml} provided in\textsuperscript{\ref{PoRbit}}, which we used for our star formation run.\\

\hspace{-0.2in}\fbox{\begin{minipage}{25em}
\&PHYSICS\_PARAMS \\
cooling=.true. \\
g\_star=1.6666D0 \\
n\_star=0.1D0 \\
eps\_star=0.0D0 \\ 
t\_star=3.0d0  (star forming timescale in Gyr) \\
T2\_star=4.0d4 \\
/
\end{minipage}}
\\

All the above parameters are described in the \textsc{ramses} manual. The \textit{t\_star} parameter is the star formation timescale in Gyr. Setting it to a finite, non-zero value activates star formation. One can add other parameters as per requirements.

\section{Random turbulence generation}
\label{sec:random}

To allow for initial turbulence and (optionally) density fluctuations, a random perturbation algorithm has been included based on the square-square subdivision \citep{Miller_1986}. It is similar to the well-known diamond-square algorithm widely used for the generation of random terrains, but provides a higher quality of randomness and fewer artefacts. The algorithm first applies random perturbations on a $2 \times 2 \times 2$ cubic array. At each subsequent step, the cube cells are subdivided into $2 \times 2 \times 2$ arrays and perturbed again, while the magnitude of the perturbation is reduced $2 \times$ (unless the user chooses a different value). Additional factors can be applied to the magnitude of each step, following a user-defined power spectrum. The resulting random noise is then multiplied with the density and/or the three velocity components to get turbulence.

The algorithm requires some additional variables to be set in the {\small{\textbf{\&Merger\_Params}}} in the \textit{namelist} file, and an extra parameter file \textit{qqm3d.par}. The extra lines in the \textit{namelist} are:\\

\hspace{-0.2in}\fbox{\begin{minipage}{25em}
    \&MERGER\_PARAMS \\
    \dots\\
    flg\_qqm3d=-1 ~~~~~~!Master switch (-1 means off)\\
    devflat\_dens=1.0 ~~~~~~!density mean unperturbed level\\
    devscal\_dens=0.1 ~~~~~~!density deviation scale\\
    devflat\_vel=1.0 ~~~~~~~!same, for velocities\\
    devscal\_vel=0.1\\
    scale\_objsize=1.0 ~~~~~~~!size of the perturbation mask
\end{minipage}} \\

The master switch controls the overall usage of the random perturbation algorithm. ``-1'' means ``off'', other modes are:
\begin{itemize}
\item 0 or 10: only density is perturbed,
\item 1: only adds absolute perturbation to velocities,
\item 2: combines modes 0 and 1,
\item 11 and 12: like 1 and 2, but with velocity perturbations scaled by the circular velocity (recommended),
\item 21 and 22: like 1 and 2, but with velocity perturbation relative to actual velocities (experimental).
\end{itemize}

The parameter file \textit{qqm3d.par} contains: \\

\hspace{-0.2in}\fbox{\begin{minipage}{25em}
**** Setup parameters for qqm4ramses ****\\
8 ~~~~2.5 ~~~~4. ~~~~1 ~~~~nsize,fsize,scalh ini,balance values\\
1 ~~~~1   ~~~~1. ~~~~~~~~~init mode, deviate (0:lin, 1:Gauss), power\\
0.0 ~~~~~~~~~~~~~~~~~~~~~~~~~ initial corner master values\\
10~~~~   0~~~~   1.0~~~~ 1.0~~ hr\_mode, stop rnd h after n iter ($<=0$:off), hreduce iter factor+power\\
309562  -1~~~~~~~~~~~~~~~~ seed, seed initialization mode

---- Corner perturbation scaling ----\\
0.2~~~~~~~~~~~~~~~~~~~~~~~~ scalh00\\
---- Feature power spectrum ----\\
0.1~~~~~~~~~~~~~~~~~~~~~~~~ scalh01\\
\dots\\
----- Corner initial values -----\\
0. ~~~~~~~~~~~~~~~~~~~~~~~~~~x01\\
\dots
\end{minipage}}\\

Only the lines most relevant for beginners are shown. Other lines are mostly experimental and should be left as they are, unless the user looks at the source code for more details about their purpose. The most relevant values for hr\_mode are:
\begin{itemize}
\item 4: uses the lines from the power spectrum block as weightings. The magnitude of the first non-zero perturbation is equal to \textit{devscal} and will be reduced by \textit{hreduce} (typically $1/2$) for each refinement level.
\item 10: uses a flat power spectrum with starting level \textit{fsize}. Non-integer values are used via an interpolation scheme.
\end{itemize}

The other hr\_modes should not be used for scientific runs. For details, see the source file \textit{qqm4ramses.f90} and its subroutine \textit{init\_qqm3d}.

\section{Extraction of data with \textit{extract\_por}}
\label{sec:extraction}

For the extraction of particle data, a tool called \textit{extract\_por} was developed by Ingo Thies and used here. Now including additional features related to star formation, \textit{extract\_por\_sfr} is available here\textsuperscript{\ref{PoRbit}}. This is a user-friendly tool that does not require much time to learn. After the tool is downloaded, it can be extracted to {\small{\textbf{/home/local}}}. Installation is done by executing:

\begin{lstlisting}[language=bash]
$ make xpordata
\end{lstlisting}

Inside \textit{extract\_por}, \textit{fmtRAMSES.par} is the parameter file where the extraction parameters can be set. It contains:\\

\hspace{-0.2in}\fbox{\begin{minipage}{25em}

`\textit{path/to/your/outputfiles}'	\\
38		 		~~Output No. \\
32				~~Number of CPU threads \\
0				COM reset \\

---- RADIAL BINNING SECTION ---- \\
10. ~~~	500		    ~~~	binning radius (in system units), nbins \\

---- Image params ---- \\
1	1	0	~~~~~~~~~~	~~flgimage \\
250	250		~~~~~~~~imagecenter \\
200	200	    ~~~~~~~~image width \\
500	500		~~~~~~~~~nbx,nby \\
1.5	1.5	   ~~~~~~~~~~~hx,hy smoothing (pixel units)
\end{minipage}} \\

Again, not all the parameters are detailed here, the file itself being very well commented. Only the parameters that might be important for a beginner are shown.

\begin{enumerate}
    \item The path should only specify where the output folders are located, not the output folders themselves. Thus, \textbf{../../output\_0001} will not be recognised. 
    \item COM reset subtracts the center of mass position and velocity. It could be used if the object of interest lies outside the field of view.
\end{enumerate}

To just extract the particle positions and velocities, only parameters until the \textit{Partial COM} section are important. After setting the parameters, execute:

\begin{lstlisting}[language=bash]
$./xpordata  
\end{lstlisting}

Based on the number of output files selected, the corresponding number of \textit{part.asc} and \textit{sfr.dat} files will be created. The \textit{part.asc} files contain data in ascii format with the following column meanings:
\begin{lstlisting}[language=bash]
1-3: position, 4-6: velocity, 7: mass,
8: particle ID
\end{lstlisting}

The \textit{sfr.dat} file contains:
\begin{lstlisting}[language=bash]
1: time interval in Myr, 2: SFR in M_Sun/Myr 
\end{lstlisting}

The extraction algorithm calculates the total stellar mass in a given snapshot, and evaluates the difference in stellar mass between two snapshots. One can resolve the SFR better by changing \textit{delta\_tout} in the \textit{namelist} file, or by extracting particle birth times.

Any tool can be used to extract and plot the results from the \textit{part.asc} files. Even \textit{extract\_por} can be used for plotting, in which case all the sections below \textit{Image\_Params} can be helpful. These sections can be used to set the projected density, resolution etc. To use \textit{extract\_por} for plotting, the following suggestions might be helpful:

\begin{enumerate}
    \item In \textit{Image\_params}, unless one has a special case like \citep{Oehm_2017}, using \textit{2:rgb} or \textit{3:rgbw} is not helpful. Set it to \textit{1:gray}. This works and one can set the required projected density. 

   \item The use of binning radius might be critical for resolution. In case of poor resolution, increase the number of bins. To increase the pixel resolution/zoom in, reduce the image field of view in \textit{fmtRAMSES.par}, i.e. reduce the bin sizes. 
   
   \item Users could set the box width equal to the simulation box size and locate the galaxy manually.
   
   \item \textit{hx and hy smoothing} smoothens the image. This can be changed based on needs. 
   
   \item All parameters below \textit{hx, hy smoothing} are not to be changed.
\end{enumerate}




Run \textit{extract\_por} and expect two output files to be produced:
\begin{enumerate}
    \item part.asc
    \item image.dat
\end{enumerate}

The \textit{image.dat} has the data required for plotting the image (particle positions). The simplest way is to use \textit{gnuplot}:

\begin{lstlisting}
$ gnuplot --> plot ``image.dat'' with image 
\end{lstlisting}


\section{Tests and publications using \bf\textsc{por}}
\label{sec:Test}

Since its development in 2015, \textsc{por} has been applied to a variety of problems. A first implementation showed that the observed dynamics in polar ring galaxies is explained naturally in MOND \citep{Lughausen_2013}. \citep{Renaud_2016}  compared Antennae-like galaxy encounters in MOND and in dark matter models, studying the evolution towards merging and the triggering of star formation in both models. The Galactic tidal streams of Sagittarius \citep{Thomas_2017} and Palomar 5 \citep{Thomas_2018} were investigated as gravitational experiments, with the latter's asymmetry interpreted as evidence for the EFE. \citep{Bilek_2018} showed that the satellite galaxy planes of the MW and M31 might arise from a past encounter between them. \citep{Wittenburg_2020} showed that exponential disc galaxies form naturally in MOND out of collapsing post-Big Bang gas clouds. \citep{Banik_2020_M33} simulated M33, finding that its long-term evolution is well understood in MOND, especially its weak bar and lack of a bulge. Their work also details some of the numerical methods, especially in \textsc{h-dice} and the \textit{extfield} patch.

\section{Conclusions}
\label{Conclusions}

\textsc{por} \citep{Lughausen_2015} is a general-purpose $N$-body and hydrodynamical solver for MOND. It is based on adapting \textsc{ramses}, whose modern version is not compatible with the \textsc{por} patch. It is recommended to use \textsc{por} from here\textsuperscript{\ref{PoRbit}}. This manual is a generic outline with which one can understand the basics required to set up, run, and analyse \textsc{por} simulations. The above-mentioned files like the \textit{namelist} and patches like \textit{staticpart} and \textit{hydro} are custom-made for a specific purpose, so care should be taken before using them for a different application. All the algorithms and tools mentioned in this guide are available here\textsuperscript{\ref{PoRbit}}, and are rated to work.

\section*{Acknowledgements}

IB is supported by an Alexander von Humboldt Foundation postdoctoral research fellowship. BF acknowledges funding from the Agence Nationale de la Recherche (ANR project ANR-18-CE31-0006 and ANR-19-CE31-0017) and from the European Research Council (ERC) under the European Union's Horizon 2020 research and innovation programme (grant agreement No. 834148). On a historical note, when PK joined the HISKP at Bonn University in 2013, financial means became available that allowed Fabian L\"ughausen to be hired as a PhD student co-supervised by PK and BF, to program the \textsc{por} patch for \textsc{ramses}, and to buy computer servers for MOND simulations. This led to the development of the \textsc{por} code \citep{Lughausen_2015}. The authors would like to thank Jan Pflamm-Altenburg and the referees for comments which helped to clarify this guide.


\bibliographystyle{unsrtnat}
\bibliography{PUG_bbl}
\end{multicols}
\end{document}